\documentclass{aastex} 
\usepackage{emulateapj5}

\usepackage{latexsym}
\shorttitle{He\,2-90 Jets}
\shortauthors{Guerrero et al.}

\begin{document}

\title{The Constant-Velocity Highly Collimated Outflows of the Planetary 
Nebula He\,2-90} 

\author{Mart\'{\i}n A.\ Guerrero\altaffilmark{1,2}, 
        Luis F. Miranda\altaffilmark{3}, 
        You-Hua Chu\altaffilmark{1,2}, \\
        M\'onica Rodr\'{\i}guez\altaffilmark{4}, and 
        Rosa M. Williams\altaffilmark{2,5,6}}
\altaffiltext{1}{Astronomy Department, University of Illinois, 
        1002 W. Green Street, Urbana, IL 61801, USA;
        mar@astro.uiuc.edu, chu@astro.uiuc.edu} 
\altaffiltext{2}{Visiting astronomer, Cerro Tololo Inter-American
         Observatory, National Optical Astronomy Observatories, 
        operated by the Association of Universities for Research 
        in Astronomy, Inc., under a cooperative agreement with the 
        National Science Foundation.}
\altaffiltext{3}{Instituto de Astrof\'{\i}sica de Andaluc\'{\i}a, CSIC, 
        Apdo. Correos 3004, E-18080 Granada, Spain; 
        lfm@iaa.es}
\altaffiltext{4}{Instituto Nacional de Astrof\'\i sica, Optica y 
        Electr\'onica, INAOE, Apdo. Postal 51 y 216, 72000 Puebla, 
        Pue., M\'exico; 
        mrodri@inaoep.mx}
\altaffiltext{5}{National Research Council Associate} 
\altaffiltext{6}{NASA's GSFC, code 662, Greenbelt, MD 20771, USA;
        rosanina@lhea1.gsfc.nasa.gov}

\begin{abstract}

We present high-dispersion echelle spectroscopic observations and a 
narrow-band [N~{\sc ii}] image of the remarkable jet-like features 
of He\,2-90.  
They are detected in the echelle spectra in the H$\alpha$ and 
[N~{\sc ii}] lines but not in other nebular lines.  
The [N~{\sc ii}]/H$\alpha$ ratio is uniformly high, $\simeq$ 1.  
The observed kinematics reveals bipolar collimated outflows in the 
jet-like features and shows that the southeast (northwest) component   
expands towards (away from) the observer at a remarkably constant 
line-of-sight velocity, 26.0$\pm$0.5 km~s$^{-1}$.
The observed expansion velocity and the opening angle of the jet-like
features are used to estimate an inclination angle of $\simeq5\arcdeg$ 
with respect to the sky plane and a space expansion velocity of $\simeq$ 
290~km~s$^{-1}$. 
The spectrum of the bright central nebula reveals a profusion of Fe
lines and extended wings of the H$\alpha$ line, similar to those seen 
in symbiotic stars and some young planetary nebulae that are presumed
to host a mass-exchanging binary system.  
If this is the case for He\,2-90, the constant velocity and direction 
of the jets require a very stable dynamic system against precession 
and warping.

\end{abstract}




\section{Introduction}

He\,2-90 is a planetary nebula (PN) originally discovered by \citet{he67}.  
It was classified as a compact PN Be star, i.e., a low-mass star in 
or evolving into the PN phase, by \citet{la98} based on spectroscopic 
observations by \citet{co93} (hereafter CdFM93).  
Recent $HST$ images of He\,2-90 in the H$\alpha$ line have revealed 
remarkably linear jet-like bipolar features consisting of six pairs 
of evenly spaced knots emanating from the bright central nebula 
\citep[hereafter SN00]{sa00}.  
Similar straight jet-like features have been observed in young stellar 
objects \citep[e.g., HH\,30,][]{bu96}, but the high N/O abundance ratio 
clearly indicates that He\,2-90 is an evolved object (CdFM93).

Collimated outflows are frequently seen in PNs \citep[][ and 
references therein]{gmc00,gcm01}, but the straight, intermittent, jet-like 
features in He\,2-90 are unique among PNs.  
To study the intriguing features in He\,2-90, we have obtained 
narrow-band images in the [N~{\sc ii}] emission line and long-slit 
echelle spectroscopic observations. 
These observations not only confirm the collimated outflow nature 
of the jet-like features, but also detect additional pairs of knots 
that were not detected by the $HST$ images and reveal new properties 
of He\,2-90. 
We report these observations and results in this paper.  

\section{Observations}

A narrow-band image of He\,2-90 in the [N~{\sc ii}] $\lambda$6584 
emission line was obtained on 2000 December 5 with the 0.9~m 
telescope at the Cerro Tololo Inter-American Observatory (CTIO).  
The central wavelength and FWHM of the [N~{\sc ii}] filter are 6584 
\AA\ and 15 {\AA}, respectively.  
The detector was the Tek 2K $\#3$ CCD with a pixel size of 24~$\mu$m.  
The image scale was $0\farcs396$ pixel$^{-1}$ and the seeing during the 
observations was $1\farcs1$, as measured from field stars in the image. 
The integration time was 300 s.  

High-dispersion spectroscopic observations of He\,2-90 were obtained 
on 2000 December 4 and 5 using the echelle spectrograph on the CTIO 
4~m telescope.  
The slit was oriented at PA $130\arcdeg$ along the jet-like features 
of He\,2-90 (see Appendix for a detailed explanation of the slit 
orientation). 
On the first night, the spectrograph was set up in the multi-order
mode to cover the spectral range from 3880 \AA\ to 7200 \AA\ with a slit 
length of $20\arcsec$. 
On the second night, the spectrograph was used in the single-order, 
long-slit mode, covering only the H$\alpha$ and [N~{\sc ii}] 
$\lambda\lambda$6548,6584 lines but with a slit length of $\sim$3$'$.
The total integration times were 300 s and 600 s for the multi-order and 
single-order observations, respectively.  
In both cases, the spectrograph was used with the 79 line~mm$^{-1}$ 
echelle grating and the long-focus red camera; the resultant
reciprocal dispersion was 3.4 \AA~mm$^{-1}$.  
The data were recorded with the SITE 2K $\#6$ CCD with a pixel size 
of 24 $\mu$m.  
This configuration provides a spatial scale of $0\farcs26$ pixel$^{-1}$ 
and a sampling of 3.7 km~s$^{-1}$~pixel$^{-1}$ along the dispersion direction.
The multi-order and single-order observations were made with slit widths
of $1\farcs0$ and $1\farcs6$, and the resultant instrumental FWHMs were 
8 km~s$^{-1}$ and 12 km~s$^{-1}$, respectively.
The angular resolution, determined by the seeing, was $\sim0\farcs9$. 
Observations of the spectrophotometric standard HR\,3454 were used
for flux calibration.

\section{Results} 

\subsection{Morphology in the [N~{\sc ii}] Narrow-band Image}

The [N~{\sc ii}] image of He\,2-90 is shown in Figure~1.  
The straight jet-like features described by SN00 in H$\alpha$ are 
clearly present in [N~{\sc ii}], and can be traced down to 
$3\arcsec$ from the center of the bright core of the nebula.
The surface brightness of the jet-like features decreases steeply with 
radial distance.  
Emission is detected up to a radial distance of $11\farcs5$, and
possibly up to $\sim18\arcsec$.  
Our spatial resolution does not allow us to clearly resolve the individual 
knots described by SN00, but the surface brightness enhancements at
$4\farcs5$, $5\farcs8$, and $8\farcs3$ from the center appear to correspond 
to SN00's knots cc$'$, dd$'$, and ff$'$.

\subsection{Spectra of the Jet-like Features} 

The echellograms of the H$\alpha$ and [N~{\sc ii}] $\lambda$6584 lines 
observed in the single-order, long-slit mode are shown in Figure~2.  
Both lines reveal a bipolar outflow structure in the jet-like features.
The velocity difference between the approaching southeast and receding 
northwest components is $\simeq52$~km~s$^{-1}$.
The observed FWHM of the [N~{\sc ii}] line in these outflows is 
typically $\simeq$16~km~s$^{-1}$, corresponding to an intrinsic
FWHM of $\simeq$10~km~s$^{-1}$ (removing the instrumental contribution
to the observed width).
The H$\alpha$ line in the outflows is wider, with an observed FWHM 
of $24-30$~km~s$^{-1}$, corresponding to an intrinsic FWHM of 
$20-27$~km~s$^{-1}$.  
The systemic velocity of the bipolar outflows and the central nebula
is V$_{\rm LSR}$ $\simeq$ $-$30~km~s$^{-1}$, in agreement with the 
value reported by CdFM93.
The radial velocity along the outflows is remarkably constant, with
variations smaller than $\pm$ 0.5~km~s$^{-1}$, i.e., less than 2\% 
of the observed velocity.  
Regardless of the inclination angle, the fractional variations of 
the expansion velocity of these ouflows ought to be the same as the 
fractional variations of the observed radial velocity, i.e., 
$\lesssim 2\%$, unless the tangential velocity does not co-vary with 
the radial velocity (an unlikely situation).  
The constant radial expansion velocity and the straight linear 
morphology of the jet-like features in He\,2-90 can only be understood 
if they are expanding with a constant space velocity.  

The [N~{\sc ii}] echellogram shows knotty structure in the jet-like
features.
The surface brightness of the [N~{\sc ii}] line along the slit is
plotted in Figure~3 to illustrate the distribution of the knots.
The counterparts of SN00's knots bb$'$ and dd$'$ are clearly seen, 
but the other knots cannot be discerned because of our poorer 
spatial resolution.
In addition, we find five more pairs of knots extending beyond 
SN00's knots ff$'$; these newly discovered knots have been labeled as  
gg$'$, hh$'$, ii$'$, jj$'$, and kk$'$ in Figures 2 and 3.  
The pairs of knots hh$'$ and jj$'$ at radial distances up to 
$11\farcs5$ and $18\farcs7$ from the center correspond to the 
positive and possible detections of emission in the [N~{\sc ii}] 
image (Fig.~1).  

The jet-like features of He\,2-90 have a uniformly high [N~{\sc ii}] 
to H$\alpha$ ratio of $\simeq$1 (Fig.~2).  
The multi-order echelle observations of He\,2-90 allow us 
to look for emission from the jet-like features along the $20\arcsec$-long 
slit at other spectral lines in which the central nebula is detected.
However, no emission in any other line is detected in this 5-minute 
echelle observation.
More specifically, no [O~{\sc iii}], [O~{\sc i}], or [S~{\sc ii}]
emission from the jet-like features is detected.  
This spectral property is very different from that of the central nebula 
which has an [O~{\sc iii}] line strength similar to [N~{\sc ii}].  
It is also different from those observed in shocks, where the 
[O~{\sc i}] and [S~{\sc ii}] lines are usually enhanced.  

Finally, we note that an additional pair of faint bipolar features 
are detected in the H$\alpha$ echellogram, but not in [N~{\sc ii}],
in Fig.~2.
This pair of bipolar features expand in the opposite direction to
that of the aforementioned prominent bipolar outflow, but with 
a smaller velocity difference ($\Delta$V $\simeq$ 42 km~s$^{-1}$) and
a shorter spatial extent ($\pm$8$\arcsec$).

\subsection{Spectrum of the Central Nebula}

The echelle observations of the central nebula reveal a wealth of 
emission lines. In addition to the H~{\sc i} Balmer, He~{\sc i}, 
[O~{\sc i}], [O~{\sc iii}], [Ar~{\sc iii}], and [N~{\sc ii}] lines 
described by CdFM93, we also detect C~{\sc ii},
Fe~{\sc ii}, [Fe~{\sc ii}], [Fe~{\sc iii}], [Fe~{\sc iv}], 
[Ni~{\sc iii}], and Si~{\sc ii} emission lines. 
Among all these species, [Fe~{\sc iii}] exhibits by far the 
strongest lines and the richest spectrum.  
The identification of these lines, and their measured strength, $F$, 
and intrinsic strength, $I$, normalized to the H$\beta$ line (=100.0) 
are listed in Table~1.  
The H$\beta$ flux is $1.8 \times 10^{-12}$ erg~cm$^{-2}$~s$^{-1}$.  
The extinction is determined from Balmer decrements using the 
H$\alpha$, H$\beta$, and H$\gamma$ lines.  
We derive an extinction $c_{{\rm H}\beta} = 2.1$ in good agreement 
with CdFM93 value.  
This amount of extinction and the extinction factor derived from 
Whitford's (1958) reddening law, $f(\lambda)$, are used to underedden 
the spectrum.  
 
The profiles of the emission lines can be grouped into three broad 
categories:  
double-peaked profiles with peak-to-peak separation $\sim 32$ 
km~s$^{-1}$ (e.g., [O~{\sc iii}], [Ar~{\sc iii}], He~{\sc i}, and 
H$\alpha$ and H$\beta$), broad single-peaked profiles (FWHM $\ga$ 
40 km~s$^{-1}$) apparently composed by two blended components (e.g., 
[N~{\sc ii}], and [Fe~{\sc iii}]), and narrow (FWHM $\la$ 30 
km~s$^{-1}$) single-peaked profiles (e.g., [S~{\sc ii}], and 
[O~{\sc i}]). 
Examples of these are shown in Figure 4.  
In all the double-peaked profiles, the red peak is brighter than the 
blue one (see Table~1), thus indicating that the brighter component 
is receding and the fainter component is approaching.  
The much more reduced peak brightness contrast of the H$\alpha$ line 
in our observations ($\sim1.4$) than in SN00's $HST$ observations 
($\sim 7$) can be attributed to our limited spatial resolution and 
coarse sampling (the angular distance between the northwest and 
southeast components is $\sim 0\farcs25$, i.e., the pixel size of our 
observations, and more than 3 times smaller than the seeing).  
Bearing this in mind, we conclude that the brighter northwest component 
is receding rather than approaching as assumed by SN00.  

Very extended wings are observed in the H$\alpha$ line (Fig.~5) 
with a full-width at zero intensity of $\pm 1500$~km~$^{-1}$ 
($\sim67$~{\AA}).  This value is larger than $\pm 1050$~km~$^{-1}$ reported 
by CdFM93. Although this difference could be due to the higher spectral 
resolution and/or sensitivity of our observations, we cannot rule out that 
the H$\alpha$ line profile has dramatically varied in the past 8 years.  
This possibility requires further investigation.  

The anomalously high ratios of [O~{\sc iii}] $\lambda4363/(\lambda4959 
+ \lambda5007)$ $\simeq$ 0.16 and [N~{\sc ii}] $\lambda5755/(\lambda6548 
+ \lambda6584)$ $\simeq$ 0.022 indicate that these lines arise from high 
density regions.
The high density is also indicated by the [S~{\sc ii}] 
$\lambda\lambda6717,6731$ doublet ratio, which is at the
high-density limit ($> 10^4$ cm$^{-3}$).
Assuming that $T_{\rm e}$ $\approx$ 15,000 K, the measured [O~{\sc iii}] 
and [N~{\sc ii}] line ratios imply $N_{\rm e}$ $\simeq 1.5\times10^5$ 
cm$^{-3}$, consistent with the value reported by CdFM93.  
A lower $T_{\rm e}$ would result in an even higher $N_{\rm e}$.   
The relative intensities of the [Fe~{\sc iii}] lines 
also indicate $N_{\rm e} = 1.6\times10^5-3.2\times10^5$ cm$^{-3}$, and 
$T_{\rm e} = 10,000-15,000$~K.

\section{Discussion}

Despite its unique morphology, He\,2-90 has spectral properties
similar to young PNs, e.g., M\,1-91, M\,1-92, M\,2-9, and IC\,4997 
\citep{ba89,go91,so94,mi96,bu98}, or symbiotic stars, e.g., RX\,Pup, 
R\,Aqr, and V1016\,Cyg \citep{su85, iv94,co99,mi99}.  
Rich Fe emission spectra from high density cores, extended wings 
in the H$\alpha$ line, extreme bipolar morphologies and collimated 
outflows are all common features of these nebulae. 
Our observations of He\,2-90 do not allow us to ascertain the presence 
or lack of TiO absorption bands, a characteristic signature of 
symbiotic stars.  
Therefore, the exact nature of He\,2-90 remains uncertain.

The kinematic properties of the jet-like features confirm their 
outflow nature and high collimation. 
The low line-of-sight expansion velocity most likely indicates 
that the outflow moves close to the plane of the sky.  
As with SN00, we also assume that the opening angle of the jet-like 
features ($\theta \sim 4\arcdeg$) is determined by the transverse 
expansion of the knots.
The profile of the [N~{\sc ii}] line can be decomposed into thermal 
and turbulent components with FWHMs of 5.7 km~s$^{-1}$ (for 10$^4$ K) 
and 8.7 km~s$^{-1}$, respectively. 
The FWHM of the turbulent component is much smaller than the FWHM of 
the thermal component of hydrogen (21.4 km~s$^{-1}$ at 10$^4$ K); 
therefore, the transverse expansion velocity ($v_{\rm t}$) of the knots 
is dominated by the sound velocity $\sim$10 km~s$^{-1}$ which is 
consistent with the FWHM of the H$\alpha$ line.  
The expansion velocity projected on the sky plane\footnote{
Note that SN00's derivation of the jet velocity contains an error.  
They used the full opening angle, instead of half of the opening 
angle, and the inclination in the relation between the transverse 
expansion velocity and the jet velocity.  
This caused the difference between their and our results.}    
is $v_{\rm sky} = v_{\rm t}/tan(\theta/2)$ $\simeq$ 286\,km~s$^{-1}$.
From this $v_{\rm sky}$ and the observed line-of-sight expansion 
velocity, 26 km~s$^{-1}$, we derive a space expansion velocity 
of $\simeq$ 287\,km~s$^{-1}$ and an inclination angle of 
$\simeq 5\arcdeg$ with respect to the sky plane.  
Notice that the uncertainties associated to the determination of the 
opening angle of the jet-like features and their transverse expansion 
velocity can introduce a large error in the values derived for the 
space expansion velocity and inclination angle.  
For the estimated values of the space expansion velocity and 
inclination angle, the expected proper motion is $\sim 
0\farcs06/d$~yr$^{-1}$, where $d$ is the distance in kpc to He\,2-90.  
It is possible to measure this proper motion with multi-epoch $HST$ 
observations.  

The large expansion velocity and high collimation of the bipolar 
features in He\,2-90 qualify them to be called ``jets".
This expansion velocity is low compared to those expected in jets 
arising from accretion disks around massive compact objects, i.e., 
neutron stars and black holes \citep{li99}.  
However, the bipolar outflows in He\,2-90 are in accord with fast 
collimated outflows in many PNs and proto-PNs, such as He\,3-1475, 
MyCn\,18, NGC\,2392, and OH\,231.8+4.2 
\citep{bob95,bry97,gbs85,rie95,san00}.  
Similarly, the high [N~{\sc ii}]/H$\alpha$ ratio and the lack of 
[O~{\sc i}] and [S~{\sc ii}] emission, while contrary to shock-excited
spectra of Herbig-Haro (HH) objects \citep{ra96}, are typical for 
FLIERs and fast outflows in PNs \citep{Hajian97,Balick98,ocon00}.  

The kinematics of the bipolar outflows shows that the southeastern 
component is approaching and the northwestern component receding. 
This disagrees with the orientation proposed by SN00 based on 
the surface brightness variation across the central nebula
and the assumption that differential obscuration makes the 
approaching side appear brighter than the receding side.
Our echelle spectrum of the central nebula shows, however, that 
the receding component is brighter than the approaching component 
(see Fig.~4), contrary to SN00's assumption.  
A similar situation is observed in OH\,231.8+4.2 of which the brighter 
lobe is receding from us \citep{san00}.  

The observations reported in this paper confirm the presence of 
highly collimated bipolar outflows in He\,2-90.  
We find that all material in the bipolar outflows expands at a 
constant velocity, with $\lesssim$ 2\% fractional variations of 
the expansion velocity.  
We have further estimated the expansion velocity and inclination of 
the outflows.  
These results, in conjunction with upcoming $HST$ imagery to study 
the proper motion of the jets ($HST$ program ID: 9102, PI: R.\ Sahai), 
will help constrain the distance to He\,2-90 and the dynamics of the 
bipolar outflows. 
We also find that the spectrum of the central nebula of He\,2-90 is 
very similar to that observed in some proto-PNs, young PNs, and 
symbiotic stars whose origin is most likely related to the evolution 
of mass-exchanging close binary systems.  
If the bipolar outflows of He\,2-90 are caused by an accretion disk 
associated with such a binary system, the periodic ejection of knots 
along the same direction at a constant velocity requires an extremely 
stabledynamics of the system against precession and warping of the 
accretion disk.  
Considering that most jets and outflows in PNs show evidence of 
precession and wobbling, the stable dynamics of He\,2-90 is truly 
unusual.

\acknowledgements
We thank Jorge Casares, Enrique P\'erez, and Hugo E.\ Schwarz for 
valuable comments on the subject, and Jos\'e Franco, Guillermo 
Garc\'{\i}a-Segura,  and J.\ Alberto L\'opez for enlightening 
discussions.  
Robert A.\ Gruendl provided assistance in reducing and analyzing 
the echelle data.  
LFM is supported partially by DGESCIC PB98-0670-C02.  
This research was partially supported by an American Astronomical Society 
Small Research Grant funded by NASA.

\appendix

Our echelle observations show that the southeast side of He\,2-90 
is blue-shifted and the northwest side red-shifted.  
Because this direction is opposite to that proposed by SN00 based 
on the surface brightness variation of the bright central nebula, 
the referees were concerned that our slit orientation was reversed.  
At the referee's request, we describe in detail in this appendix 
how our slit orientation was determined and verified.

The echelle spectrograph on the CTIO 4 m telescope can be rotated 
at the Cassegrain mount, and the slit position is recorded as the 
``rotation angle" which is related to the position angle (measured 
counterclockwisely from the north) by rotation angle = position 
angle + 180$^\circ$.  
In the data array, the vertical axis is spatial and the horizontal 
axis is spectral.
The direction toward the position angle is at the bottom of the 
echellogram.
For example, a EW slit has a position angle of 90$^\circ$, a rotation 
angle of 270$^\circ$, and east is at the bottom of the echellogram;
a NS slit has a position angle of 0$^\circ$, a rotation angle of 
180$^\circ$, and north is at the bottom of the echellogram.
The relationship between the position angle and the orientation in
the echellogram has been verified by observations of supernova 
remnants in the Magellanic Clouds.  
The surface brightness of these supernova remnants has no symmetry, 
and hence offers unambiguous verification.

The slit position we used for the observations of He\,2-90 was along 
the position angle 130$^\circ$, corresponding to a rotation angle of 
310$^\circ$, thus the southeast side should be at the bottom and the 
northwest side at the top of the echellogram.  
The echellogram shows that the southeast side of the outflow is 
approaching and the northwest side of the outflow is receding.

Another indirect confirmation of the slit orientation is the
spectrum of the central nebula.  
The $HST$ H$\alpha$ image of the central nebula shows that the 
southeast lobe is much brighter than the northwest lobe, which formed 
the basis of SN00's assumption that the northwest lobe is obscured 
and is receding.  
Our spectrum of the central nebula shows that the receding component
is brighter than the approaching component.  
The brightness contrast between the receding component and the 
approaching component is not as dramatic as the contrast between the 
southeast and northwest lobes of the central nebula in the $HST$ image 
because the echelle spectrum is extracted from a large aperture 
(1$''$$\times$2\farcs5) that includes both lobes.  
The inclusion of extended emission has diluted the contrast in the 
very central region.  
While our echelle observation of the central nebula does not have any 
spatial information, the comparison of intensity variation in the 
velocity profile and the surface brightness variation in the $HST$ 
image suggests that the bright northwest side is receding and the faint 
southeast side is approaching.
This is in full agreement with the orientation of the outflows.

\newpage


\begin{deluxetable}{llrrc|llrrc}
\tablenum{1}
\tablewidth{43pc}
\tablecaption{Measured and Intrinsic Line Intensity Ratios for He\,2-90} 
\tablehead{
\multicolumn{1}{c}{$\lambda_{\rm obs}$} & 
\multicolumn{1}{c}{Line} & 
\multicolumn{1}{c}{$F$} & 
\multicolumn{1}{c}{$f(\lambda)$} & 
\multicolumn{1}{c}{$I$} & 
\multicolumn{1}{c}{$\lambda_{\rm obs}$} & 
\multicolumn{1}{c}{Line} & 
\multicolumn{1}{c}{$F$} & 
\multicolumn{1}{c}{$f(\lambda)$} & 
\multicolumn{1}{c}{$I$} \\
\multicolumn{1}{c}{[\AA]} & 
\multicolumn{1}{c}{} & 
\multicolumn{1}{c}{} & 
\multicolumn{1}{c}{} & 
\multicolumn{1}{c}{} & 
\multicolumn{1}{c}{[\AA]} & 
\multicolumn{1}{c}{} & 
\multicolumn{1}{c}{} & 
\multicolumn{1}{c}{} & 
\multicolumn{1}{c}{} 
}

\startdata 
4101.3  & H~{\sc i}      $\lambda$4101.7   &    10.0  &   0.172 &   24.0    & 5875.0  & He~{\sc i}     $\lambda$5875.7   &    47.2  & --0.126 &   15.5     \nl  
4340.05 & H~{\sc i}      $\lambda$4340.5   &    25.3  &   0.129 &   48.6    & 5874.67 & He~{\sc i}     $\lambda$5875.7 blue &    17.1  & --0.126 &    5.6  \nl  
4343.0  & [Fe~{\sc ii}]  $\lambda$4343.8 ? &     3.1  &   0.128 &    5.9    & 5875.29 & He~{\sc i}     $\lambda$5875.7 red  &    30.1  & --0.126 &    9.9  \nl  
4362.7  & [O~{\sc iii}]  $\lambda$4363.2   &     3.3  &   0.124 &    6.2    & 5957.1  & Si~{\sc ii}    $\lambda$5957.6   &     0.6  & --0.230 &    0.19    \nl  
4471.1  & He~{\sc i}     $\lambda$4471.5   &     3.0  &   0.095 &    4.9    & 5978.4  & Si~{\sc ii}    $\lambda$5978.9   &     1.5  & --0.234 &    0.5     \nl  
4657.7  & [Fe~{\sc iii}] $\lambda$4658.1   &     3.5  &   0.049 &    4.5    & 5999.6  & [Ni~{\sc iii}] $\lambda$6000.2   &     0.6  & --0.237 &    0.18    \nl  
4701.1  & [Fe~{\sc iii}] $\lambda$4701.6   &     1.8  &   0.038 &    2.2    & 6248.4  & Fe~{\sc ii}    $\lambda$6247.6 ? &     0.3  & --0.277 &    0.07    \nl  
4769.2  & [Fe~{\sc iii}] $\lambda$4769.4   &     0.6  &   0.024 &    0.7    & 6346.5  & Si~{\sc ii}    $\lambda$6347.1   &     1.8  & --0.292 &    0.4     \nl  
4860.9  & H~{\sc i}      $\lambda$4861.33  &   100.0  &         &  100.0    & 6363.2  & [O I]          $\lambda$6363.8   &     1.3  & --0.294 &    0.3     \nl  
4860.54 & H~{\sc i}      $\lambda$4861.33 blue &    40.4  &         &  40.4 & 6370.7  & Si~{\sc ii}    $\lambda$6371.4   &     0.9  & --0.295 &    0.20    \nl    
4861.05 & H~{\sc i}      $\lambda$4861.33 red  &    59.6  &         &  59.6 & 6383.8  & Fe~{\sc ii}    $\lambda$6383.7   &     1.0  & --0.297 &    0.22    \nl    
4921.5  & He~{\sc i}     $\lambda$4921.9   &     1.8  & --0.014 &    1.7    & 6401.0  & [Ni~{\sc iii}] $\lambda$6401.5   &     0.3  & --0.300 &    0.06    \nl  
4958.4  & [O~{\sc iii}]  $\lambda$4958.9   &    68.3  & --0.023 &   62.0    & 6533.0  & [Ni~{\sc iii}] $\lambda$6533.9   &     1.0  & --0.318 &    0.19    \nl  
4958.10 & [O~{\sc iii}]  $\lambda$4958.9 blue &    29.3  & --0.023 &   26.6 & 6547.3  & [N~{\sc ii}]   $\lambda$6548.0   &    55.6  & --0.321 &   10.7     \nl  
4958.70 & [O~{\sc iii}]  $\lambda$4958.9 red  &    39.0  & --0.023 &   35.4 & 6562.14 & H~{\sc i}      $\lambda$6562.85   &  1523.4  & --0.323 &  284.7     \nl 
5006.35 & [O~{\sc iii}]  $\lambda$5006.84   &   254.0  & --0.034 &  219.7   & 6561.66 & H~{\sc i}      $\lambda$6562.85 blue &   555.5  & --0.323 &  103.8     \nl
5006.01 & [O~{\sc iii}]  $\lambda$5006.84 blue &   113.9  & --0.034 &  98.5 & 6562.43 & H~{\sc i}      $\lambda$6562.85 red  &   777.7  & --0.323 &  145.3     \nl
5006.62 & [O~{\sc iii}]  $\lambda$5006.84 red  &   140.1  & --0.034 & 121.2 & 6577.3  & C~{\sc ii}     $\lambda$6578.0   &     0.8  & --0.325 &    0.15    \nl  
5010.8  & [Fe~{\sc iii}] $\lambda$5011.3   &     2.3  & --0.035 &    2.0    & 6582.7  & [N~{\sc ii}]   $\lambda$6583.37  &   171.7  & --0.326 &   33.2     \nl  
5015.3  & He~{\sc i}     $\lambda$5015.7   &     4.0  & --0.036 &    3.5    & 6515.8  & [S~{\sc ii}]   $\lambda$6716.5   &     0.83 & --0.343 &    0.15    \nl  
5055.4  & Si~{\sc ii}    $\lambda$5056.0   &     0.5  & --0.045 &    0.4    & 6730.1  & [S~{\sc ii}]   $\lambda$6730.8   &     1.91 & --0.345 &    0.33    \nl  
5270.1  & [Fe~{\sc iii}] $\lambda$5270.4   &     3.3  & --0.089 &    2.1:   & 6996.3  & [Fe~{\sc iv}]  $\lambda$6997.1   &     1.0  & --0.375 &    0.15    \nl  
5316.2  & Fe~{\sc ii}    $\lambda$5316.6   &     0.7  & --0.099 &    0.4    & 7001.4  & O~{\sc i}      $\lambda$7001.7   &     0.4  & --0.375 &    0.06    \nl  
5411.8  & [Fe~{\sc iii}] $\lambda$5412.0   &     0.4  & --0.118 &    0.2    & 7064.5  & He~{\sc i}     $\lambda$7065.3   &    35.2  & --0.383 &    5.1     \nl  
5517.5  & [Cl~{\sc iii}] $\lambda$5517.7   &     0.44 & --0.139 &    0.21   & 7135.1  & [Ar~{\sc iii}] $\lambda$7135.8   &    123.2 & --0.391 &   17.0     \nl  
5537.1  & [Cl~{\sc iii}] $\lambda$5537.9   &     0.50 & --0.143 &    0.24   & 7134.69 & [Ar~{\sc iii}] $\lambda$7135.8 blue &    54.9  & --0.391 &    7.6  \nl  
5551.2  & N~{\sc ii}     $\lambda$5551.9   &     0.3  & --0.146 &    0.15   & 7135.46 & [Ar~{\sc iii}] $\lambda$7135.8 red  &    68.3  & --0.391 &    9.4  \nl  
5754.0  & [N~{\sc ii}]   $\lambda$5754.6   &    19.8  & --0.191 &    7.2    & 7154.5  & [Fe~{\sc ii}]  $\lambda$7155.1   &     0.8  & --0.392 &    0.11    \nl  
	&				   &	      &		&           & 7171.3  & [Fe~{\sc ii}]  $\lambda$7172.0   &     0.6  & --0.394 &    0.08    \nl

\enddata
\end{deluxetable}

\clearpage


\newpage

\begin{figure}
\epsscale{0.6}
\centerline{\plotone{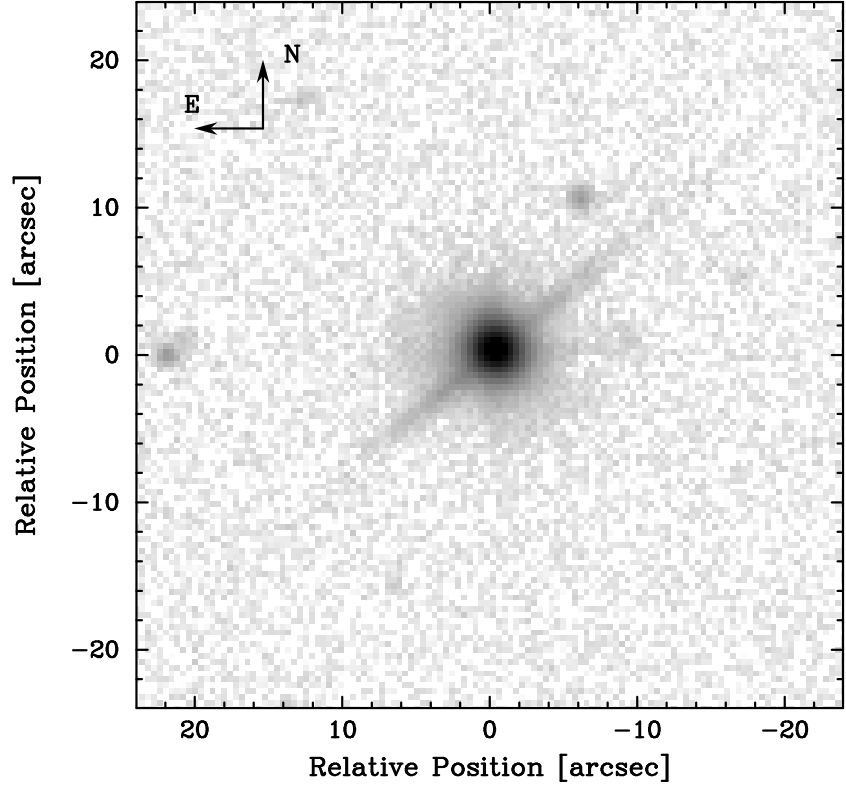}}
\caption{[N~{\sc ii}] image of He\,2-90, on a logarithmic intensity scale. 
} 
\end{figure}

\begin{figure}
\epsscale{0.9}
\centerline{\plotone{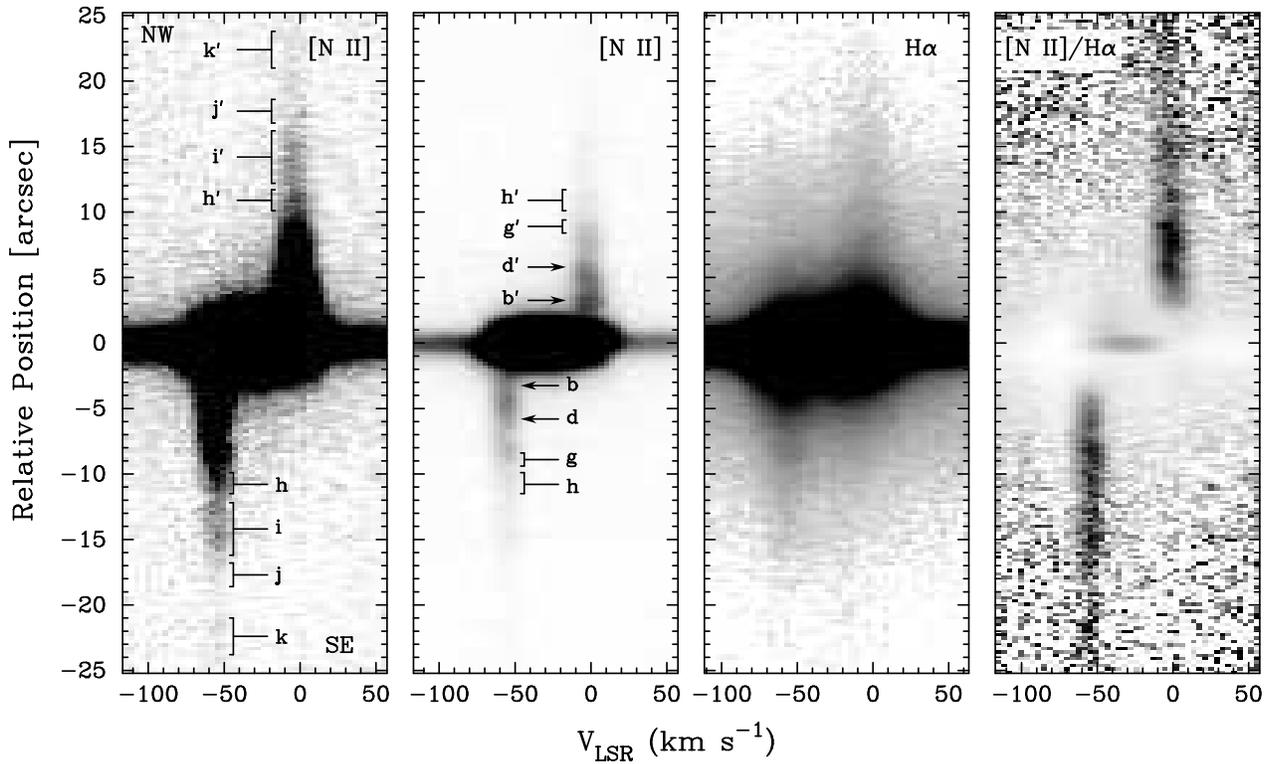}}
\caption{
[N~{\sc ii}] $\lambda$6584 and H$\alpha$ echellograms, and 
[N~{\sc ii}]/H$\alpha$ ratio map at PA = $130\arcdeg$ along the 
jet-like features of He\,2-90.  
The images are displayed on a linear scale.  
The [N~{\sc ii}] echellogram is presented at two different intensity 
levels to show both the bright and faint features.  
In the [N~{\sc ii}]/H$\alpha$ ratio map, levels are from 0 (white) to 
1.5 (black).  
The position of the pairs of knots bb$^\prime$ and dd$^\prime$ 
discovered by SN00 are marked by arrows on the [N~{\sc ii}] 
echellograms; 
the position and extension of the new pairs of knots from gg$^\prime$ 
to kk$^\prime$ are also labeled.  
}
\end{figure}

\begin{figure}
\epsscale{0.6}
\centerline{\plotone{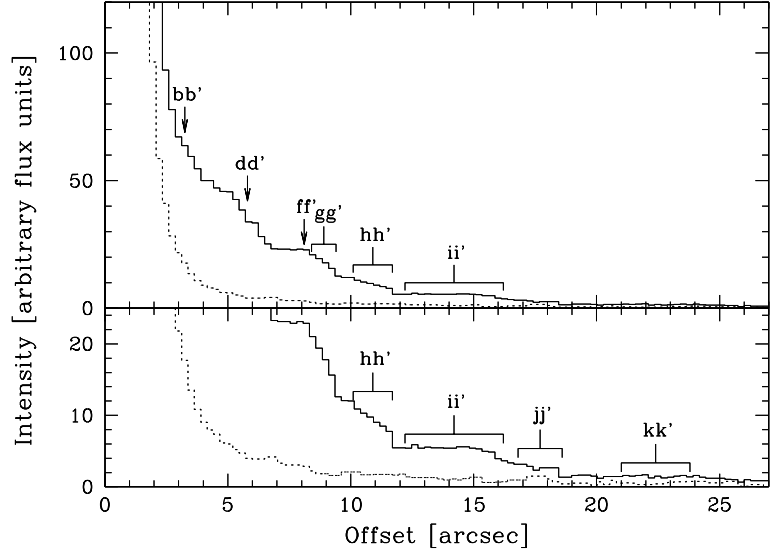}}
\caption{Spatial profile of the surface brightness of the [N~{\sc ii}] 
$\lambda$6584 emission line in the jet-like features of He\,2-90.  
The northwest and southeast components have been added to improve the 
S/N ratio.  
The different pairs of knots (see Fig.~2) are labeled.  
The dotted histogram corresponds to the brightness distribution of the 
central nebula and has been plotted to highlight the spatial extension 
and intensity of the jet-like features in He\,2-90.  }
\end{figure}

\begin{figure}
\centerline{\plotone{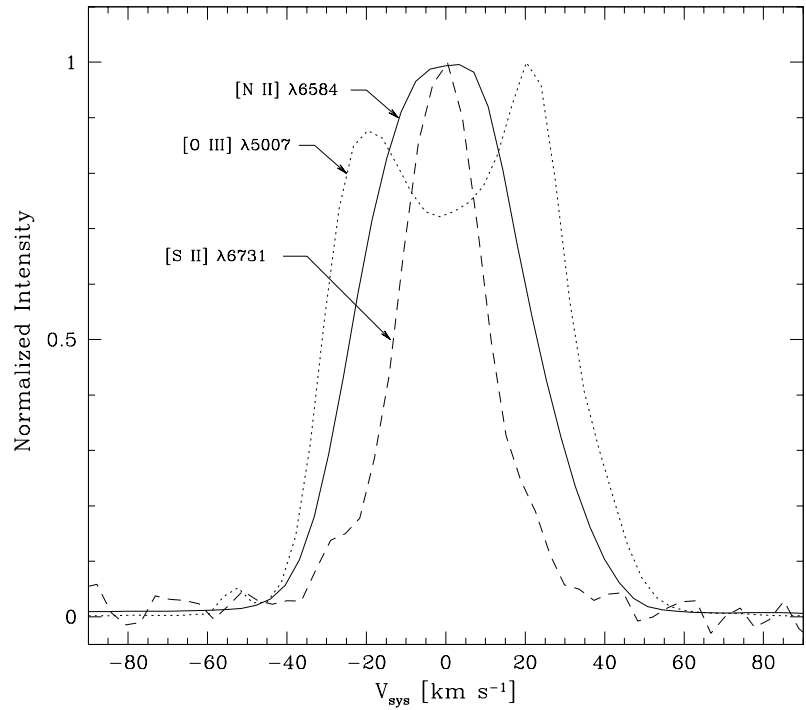}}
\caption{
Velocity profile of the [N~{\sc ii}] $\lambda$6584 (solid line), 
[O~{\sc iii}] $\lambda$5007 (dotted line), and [S~{\sc ii}] 
$\lambda$6731 (dashed line) emission lines at the bright core of 
He\,2-90.  
The intensity of each line has been normalized to its maximum value.  
The different line profile shapes and widths are noticeable.  }
\end{figure}

\begin{figure}
\centerline{\plotone{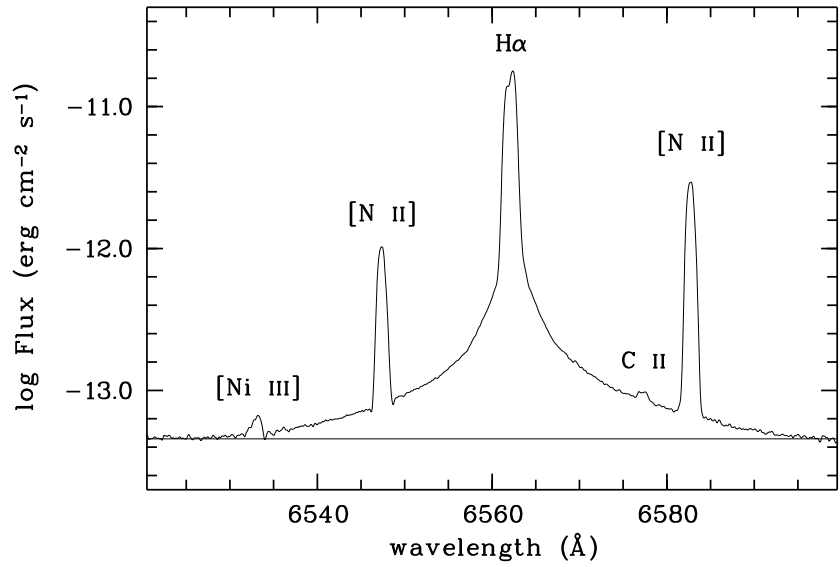}}
\caption{
Line profile of the H$\alpha$ line from the bright core of He\,2-90.  
The broad wings extend $\pm 1500$ km~s$^{-1}$.  
Some relevant emission lines are labeled.  }
\end{figure}


\begin{thebibliography}{}

\bibitem[Balick(1989)]{ba89} 
Balick, B.\ 
1989, \aj, 97, 476 

\bibitem[Balick et al.(1998)]{Balick98} Balick, B., Alexander, J., 
Hajian, A.\ R., Terzian, Y., Perinotto, M., \& Patriarchi, P.\ 
1998, \aj, 116, 360 

\bibitem[Bobrowsky et al.(1995)]{bob95} 
Bobrowsky, M., Zijlstra, A.\ A., Grebel, E.\ K., Tinney, C.\ G., Te 
Lintel Hekkert, P., van de Steene, G.\ C., Likkel, L., \& Bedding, T.\ 
R.\ 
1995, \apjl, 446, L89 

\bibitem[Bryce et al.(1997)]{bry97} 
Bryce, M., Lopez, J.\ A., Holloway, A.\ J., \& Meaburn, J.\ 1997, \apjl, 
487, L161 

\bibitem[Bujarrabal et al.(1998)]{bu98} 
Bujarrabal, V., Alcolea, J., Sahai, R., Zamorano, J., \& Zijlstra, A.\ 
A.\ 
1998, \aap, 331, 361 

\bibitem[Burrows et al.(1996)]{bu96} 
Burrows, C.\ J., et al.\ 
1996, \apjs, 473, 437

\bibitem[Corradi et al.(1999)]{co99} 
Corradi, R.\ L.\ M., Ferrer, O.\ E., Schwarz, H.\ E., Brandi, E., \& 
Garc\'{\i}a, L.\ 
1999a, \aap, 348, 978 

\bibitem[Costa, de Freitas-Pacheco, \& Maciel(1993)]{co93} 
Costa, R.\ D.\ D., de Freitas-Pacheco, J.\ A., \& Maciel, W.\ J.\ 
1993, \aap, 276, 184 (CdFM93)

\bibitem[Gieseking, Becker, \& Solf(1985)]{gbs85} 
Gieseking, F., Becker, I., \& Solf, J.\ 1985, \apjl, 295, L17 

\bibitem[Goodrich(1991)]{go91} 
Goodrich, R.\ W.\ 
1991, \apj, 366, 163 

\bibitem[Gon{\c c}alves, Corradi, \& Mampaso(2001)]{gcm01} 
Gon{\c c}alves, D.\ R., Corradi, R.\ L.\ M., \& Mampaso, A.\ 2001, \apj, 
547, 302 

\bibitem[Guerrero, Miranda, \& Chu(2000)]{gmc00} 
Guerrero, M.\ A., Miranda, L.\ F., \& Chu, Y.-H.\ 
2000, Ionized Gaseous Nebulae.\ Mexico City  November 21--24,  2000, 
in press

\bibitem[Hajian et al.(1997)]{Hajian97}
Hajian, A.\ R., Balick, B., Terzian, Y., Perinotto, M.\
1997, \apj, 487, 304

\bibitem[Henize(1967)]{he67} 
Henize, K.\ G.\ 
1967, \apjs, 14, 125 

\bibitem[Ivison, Bode, \& Meaburn(1994)]{iv94} 
Ivison, R.\ J., Bode, M.\ F., \& Meaburn, J.\ 
1994, \aaps, 103, 201 

\bibitem[Lamers et al.(1998)]{la98} 
Lamers, H.\ J.\ G.\ L.\ M., Zickgraf, F., de Winter, D., Houziaux, L., 
\& Zorec, J.\ 
1998, \aap, 340, 117 

\bibitem[Livio(1999)]{li99} 
Livio, M.\ 1999, \physrep, 311, 225 

\bibitem[Mikolajewska et al.(1999)]{mi99} 
Mikolajewska, J., Brandi, E., Hack, W., Whitelock, P.\ A., Barba, R., 
Garc\'{\i}a, L., \& Marang, F.\ 
1999, \mnras, 305, 190 

\bibitem[Miranda, Torrelles, \& Eiroa(1996)]{mi96} 
Miranda, L.\ F., Torrelles, J.\ M., \& Eiroa, C.\ 
1996, \apjl, 461, L111 

\bibitem[O'Connor et al.(2000)]{ocon00} 
O'Connor, J.\ A., Redman, M.\ P., Holloway, A.\ J., Bryce, M., 
L{\'o}pez, J.\ A., \& Meaburn, J.\ 
2000, \apj, 531, 336 

\bibitem[Raga, B{\"o}hm, \& Cant{\'o}(1996)]{ra96} 
Raga, A.\ C., B{\"o}hm, K.-H., \& Cant{\'o}, J.\ 1996, 
1996, Revista Mexicana de Astronom\'{\i}a y Astrof\'{\i}sica, 32, 161 

\bibitem[Riera et al.(1995)]{rie95} 
Riera, A., Garc\'{\i}a-Lario, P., Manchado, A., Pottasch, S.\ R., \& 
Raga, A.\ C.\ 
1995, \aap, 302, 137 

\bibitem[Sahai \& Nyman(2000)]{sa00} 
Sahai, R.\ \& Nyman, L.-{\AA}\
2000, \apjl, 538, L145 (SN00)

\bibitem[S{\' a}nchez Contreras et al.(2000)]{san00} 
S{\' a}nchez Contreras, C., Bujarrabal, V., Miranda, L.\ F., \& 
Fern{\' a}ndez-Figueroa, M.\ J.\ 2000, \aap, 355, 1103 

\bibitem[Solf(1994)]{so94} Solf, J.\ 
1994, \aap, 282, 567 

\bibitem[Solf \& Ulrich(1985)]{su85} 
Solf, J.\ \& Ulrich, H.\ 
1985, \aap, 148, 274 

\bibitem[Whitford(1958)]{w58}
Whitford, A.\ E.\ 1958, \aj, 63, 201

\end{thebibliography}
\end{document}